\documentclass[sigconf]{acmart}
\AtBeginDocument{%
  \providecommand\BibTeX{{%
    \normalfont B\kern-0.5em{\scshape i\kern-0.25em b}\kern-0.8em\TeX}}}

\copyrightyear{2020}
\acmYear{2020}
\setcopyright{usgovmixed}\acmConference[WiseML '20]{2nd ACM Workshop on
Wireless Security and Machine Learning}{July 13, 2020}{Linz (Virtual Event), Austria}
\acmBooktitle{2nd ACM Workshop on Wireless Security and Machine Learning (WiseML
'20), July 13, 2020, Linz (Virtual Event), Austria}
\acmPrice{15.00}
\acmDOI{10.1145/3395352.3402623}
\acmISBN{978-1-4503-8007-2/20/07}

\usepackage{float} 
\usepackage[ruled,vlined]{algorithm2e}

\begin{document}

\title{Algorithm Selection Framework for Cyber Attack Detection}


\author{Marc Chal\'{e}}
\affiliation{%
  \institution{Air Force Institute of Technology}
  \streetaddress{2950 Hobson Way}
  \city{Wright-Patterson AFB, Ohio}
  \country{USA}}
\email{marc.chale@afit.edu}

\author{Nathaniel D. Bastian}
\affiliation{%
  \institution{Army Cyber Institute}
  \streetaddress{2101 New South Post Road}
  \city{West Point, New York}
  \country{USA}}
\email{nathaniel.bastian@westpoint.edu}

\author{Jeffery Weir}
\affiliation{%
  \institution{Air Force Institute of Technology}
  \streetaddress{2950 Hobson Way}
  \city{Wright-Patterson AFB, Ohio}
  \country{USA}}
\email{jeffery.weir@afit.edu}

\renewcommand{\shortauthors}{Chal\'{e}, Bastian \& Weir}

\begin{abstract}
The number of cyber threats against both wired and wireless computer systems and other components of the Internet of Things continues to increase annually. In this work, an algorithm selection framework is employed on the NSL-KDD data set and a novel paradigm of machine learning taxonomy is presented. The framework uses a combination of user input and meta-features to select the best algorithm to detect cyber attacks on a network. Performance is compared between a rule-of-thumb strategy and a meta-learning strategy. The framework removes the conjecture of the common trial-and-error algorithm selection method. The framework recommends five algorithms from the taxonomy. Both strategies recommend a high-performing algorithm, though not the best performing. The work demonstrates the close connectedness between algorithm selection and the taxonomy for which it is premised.

\end{abstract}




\maketitle

\section{Introduction}

People, organizations and communities rely on the Internet of Things (IoT) to aid in almost any conceivable task that was previously performed manually. As technology advances, components of IoT have progressed into the wireless domain \cite{Kasongo_wirelss_IDS}. These emerging systems are susceptible to attack by malicious actors wishing to degrade the system or steal proprietary information \cite{debar1999towards}. Intrusion detection systems (IDS) are central to maintaining the security of modern computer networks from malicious actors \cite{maxwell2019intelligent}.  IDS have been successfully demonstrated in both the wired and wireless domain of IoT \cite{Kasongo_wirelss_IDS}. The task assigned to an IDS is to \textit{classify} network traffic as malicious or normal. Numerous studies \cite{Kasongo_wirelss_IDS} have explored meta-models to detect malicious behavior in computer networks. Maxwell et al. \cite{maxwell2019intelligent} further focused on intelligent cybersecurity feature engineering for various meta-models.

Learning algorithms may be used to formulate a meta-model. Selection of the best machine learning (ML) algorithm, including hyper-parameters, for a particular problem instance is a difficult and time-consuming task \cite{Smith}. Cui et al. \cite{cui2014accuracy} has confirmed conclusions of \cite{simpson1997use} and \cite{wang2007review} that meta-models' performance varies among problem types and problem instances. Wolpert et al. \cite{Wolpert2001} uses \textit{The Extended Bayesian Formalism} to show that given a set of learning algorithms and problems, each algorithm will outperform the others for some (equally sized) subset of problems. This phenomena has driven researchers to a trial-and-error strategy of identifying the best meta-model for a given problem. The preferred meta-model is selected by comparison of model performance metrics such as accuracy \cite{Cui_meta}. Unfortunately, the computational run time and human investment required to select a learning algorithm by trial-and-error is generally prohibitive of finding the optimal choice. 

This paper aims to advance the IDS body of knowledge by incorporating recent work in algorithm selection. Accordingly, an \textit{algorithm selection framework} is introduced. The algorithm selection framework leverages a taxonomy of ML algorithms. The framework narrows down the list of applicable algorithms based on problem characterization. Two strategies are presented to select the most preferred algorithm: \textit{rules-of-thumb} and \textit{meta-learner.} If successful, the algorithm selection framework promotes high-performance results of the IDS and assuages the computational cost of performing multiple ML algorithms.

This paper includes Related Works in Section 2. The Methodology is presented in Section 3. Section 4 contains the Results and Section 5 is the Conclusion.

\section{Related Works}

\subsection{Intrusion Detection}

As in any other domain, criminals and adversaries seek to inflict harm by exploiting weaknesses in cybersecurity systems. The rate of system intrusion incidents is increasing annually \cite{revathi2013detailed}. Landwehr et al. \cite{landwehr1994taxonomy} provides a taxonomy of all known flaws in computer systems. Special attention is provided to the category of flaws that allow exploitation by malicious actors. Commonly, a malicious actor, or their code, appears benign to a computer security system for a long enough time to exploit information or degrade the attacked system. The Trojan horse is among the most prevalent categories of a malicious attack on computer systems. It is characterized as a code that appears to provide a useful service but instead steals information or degrades the targeted system. A Trojan horse containing self replicating code is known as a virus. A trapdoor is a malicious attack in which an actor covertly modifies a system in such a way that they are permitted undetected access. Finally, a time bomb is an attack that accesses a system covertly and lies dormant until a detonation time. Upon detonation, the time bomb will inflict damage to the system either by disrupting service or destroying information.

Intrusion detection systems are a layer of network security that tracks activity patterns in a computer system to detect malicious actors before they can inflict harm. Debar et al. \cite{debar1999towards} describes efforts as early as 1981 and Sobirey \cite{Sobirey} maintains a repository of prominent IDS projects.  According to Debar et al. \cite{debar1999towards}, the success of these systems has spawned a commercial market of IDS software including brands such as Sysco Systems, Haystack Labs, Secure Networks, among others. Typically, the IDS employs a detector module that monitors system status. The detector catalogues patterns of both normal and malicious activity in a database. The detector also monitors patterns in the current system configuration. Further, the detector provides an audit of events occurring within the system. The detector leverages these data channels to generate an alarm for suspicious activity and countermeasures if necessary. An IDS is evaluated by its \textit{accuracy} of attack detection (false positive), \textit{completeness} to detect all threats (false negatives) and \textit{performance} to detect threats quickly.

NSL-KDD is a publicly available benchmark data set of network activity. NSL-KDD improves on several flaws of the well-known KDD Cup `99 benchmark data set. Most notably, NSL-KDD has rectified the 78\% and 75\% duplicate records in training and test sets, respectively \cite{revathi2013detailed}. Four classes of attacks are recorded in the data set. \textit{Denial of service} attacks bombard a network with an overwhelming quantity of data such that the computing resources are exhausted. As a result, the system cannot fulfil any legitimate computing processes. \textit{User to Root} attackers enter the network disguised as a legitimate user but seek security vulnerabilities which grant them elevated system privileges. A \textit{remote to user} attack is performed by sending data to a private network and identifying insecure access points for exploitation. \textit{Probing} is the attack technique by which the assailant studies an accessible system for vulnerabilities which will be exploited at a later time  \cite{paliwal2012denial}.

Maxwell et al. \cite{maxwell2019intelligent} and Viegas \cite{Viegas_Energy_Efficient_IDS} describe IDS tools that incorporate ML models. Unfortunately, raw network traffic data is not a suitable input for building accurate and efficient ML models. Instead, the data must be transformed as a set of vectors representing the raw data. The process of constructing such vectors is known as feature engineering, which is a non-trivial task that requires both domain knowledge and mastery of ML to capture all available information in the model. It is shown experimentally that varying the feature engineering strategy does affect classification accuracy of the IDS but no single feature is known to be superior to others.

Kasongo \cite{Kasongo_wirelss_IDS} explores IDS procedures catered for the wireless domain. The UNSW-NB15 data set was selected to derive both the training and test data sets. A wrapper-based feature extractor generated many feature vectors for comparison from a full set of 42 features. The experiment was performed for both binary and multi-class classification in which the type of attack was predicted.  Candidate  algorithms included decision Trees, Random Forest, Na\"ive Bayes, K-Nearest Neighbor, Support Vector Machines, and Feed-forward Artificial Neural Networks (FFANN). The optimal feature set consisted of 26 columns. The FFANN reflected the best classification accuracy on the full data set with 87.10\% binary and 77.16\% multi-class. Random Forest, Decision Tree, and Support Vector Machine were close behind. When the feature set and neural network hyper-parameters were optimized, the classification accuracy of the FFANN improved to 99.66\% and 99.77\% for binary and multi-class classification, respectively. 

\subsection{Algorithm Selection Problem}

\begin{figure}
    \centering
    \includegraphics[width = 8.5cm]{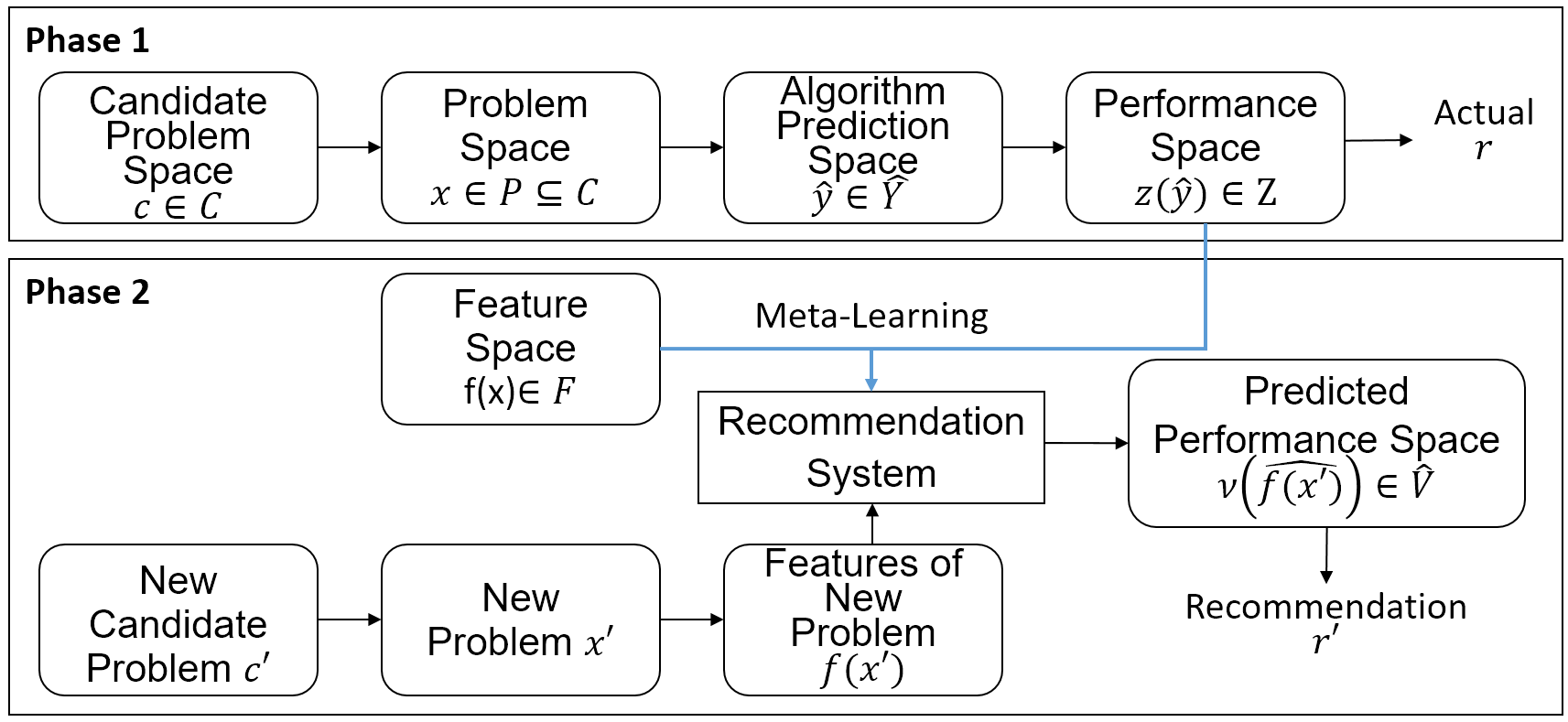}
    \caption{The meta-learner version of Rice's framework \cite{woods_thesis}\label{fig:woods_meta_framework}.}
 \end{figure}

Rice's algorithm selection framework was presented in 1976 \cite{Rice1974}. The framework is performed by employing all algorithms under consideration on all problems in a problem set. One or more performance metrics are chosen, and the performance of each algorithm on each problem is reported. Upon completion of the process, the preferred algorithm for each problem is taken as the one with the best performance metrics \cite{Rice1974}. Woods \cite{woods_thesis} presents a modern depiction of Rice's framework as phase 1 in Figure \ref{fig:woods_meta_framework}.

The classic approach of learning algorithms is known as base learning. That is an ML algorithm which builds a data-driven model for a specific application \cite{Cui_meta}. Meta-learning, however, is an approach introduced by \cite{utgoff1986shift} which algorithms learn on the learning process itself. A meta-learning algorithm extracts meta-features \( f(x) \in \) space \(F\) from a problem \textit{x} \(\in\)  problem space \(P \). The meta-model is trained to recommend the best-known base learning algorithm \( a \in A \) to solve \(x\). Works such as \cite{rendell1987layered} and \cite{brazdil1994characterizing} further contribute to the theory of meta-learning recommendation systems \cite{Cui_meta}.

In 2014, Smith \cite{Smith2014} proposes the concept of applying meta-learning to Rice's model. It was not until 2016, however, that \cite{Cui_meta} implemented the concept.  Figure \ref{fig:woods_meta_framework} demonstrates that Cui et al. \cite{Cui_meta} trained a meta-learning model to correlate problem features to algorithm performance and that the trained model could be used to recommend the algorithm for unobserved problems within Rice's framework.  The meta-learner correctly recommended the best algorithm in 91\% of test problems. Further, it demonstrated that time to perform algorithm selection could be reduced from minutes to seconds compared to trial-and-error techniques \cite{Cui_meta}. Follow on studies by \cite{woods_thesis} and \cite{williams_thesis} expanded on this work by exploring various meta-features and meta-learner response metrics.


\section{Methodology}

The assigned task for an IDS is to classify network traffic records as normal or malicious. This task is investigated from the broader perch of the algorithm selection problem.  Figure \ref{analysis_process} shows that within the analysis process, three factors drive the analytical approach and analytical technique selection. They are the input to the algorithm selection framework.

{\begin{figure}[h!]
 \begin{center}
    \includegraphics[width=8.5cm]{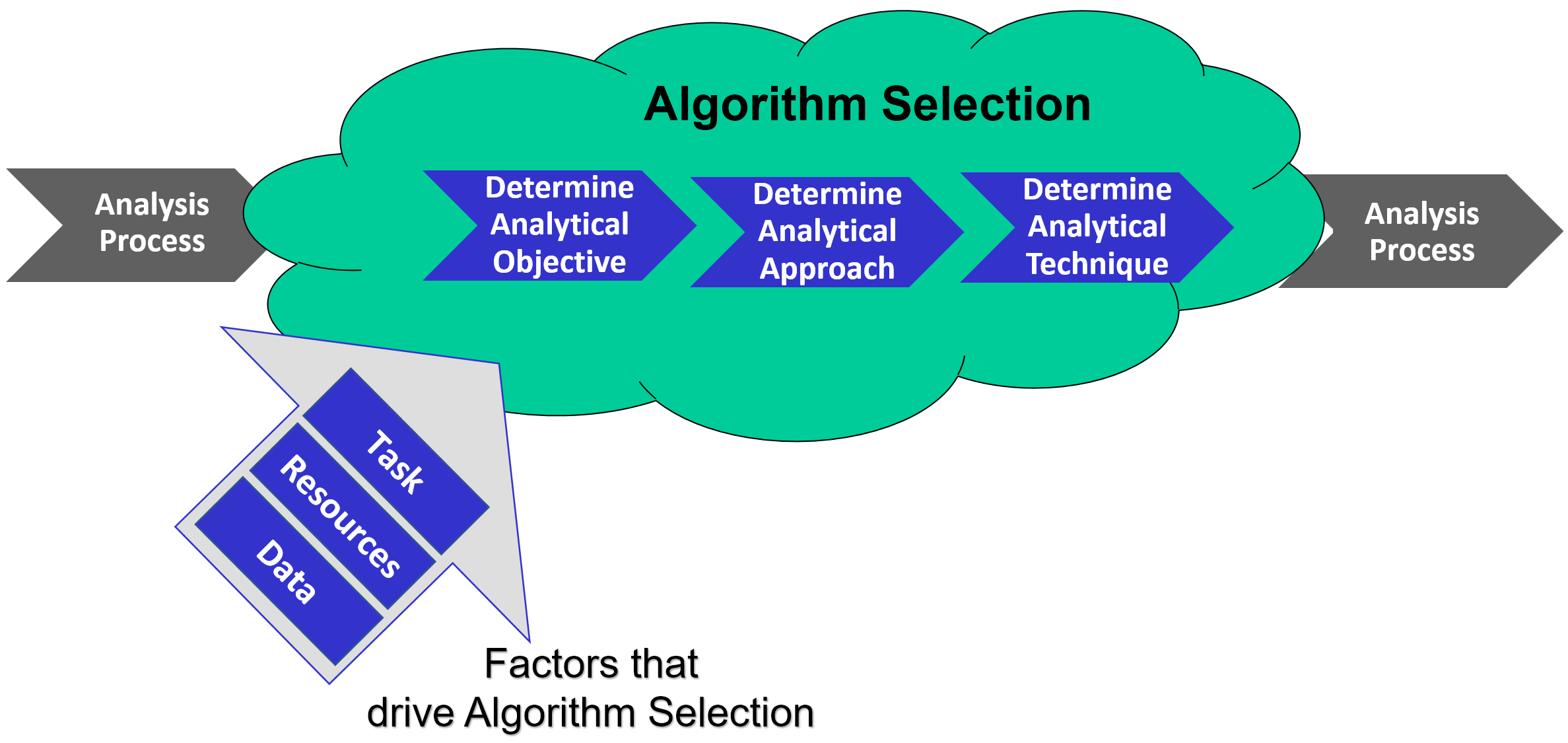}
     \caption{The factors identified are superimposed with the stages of algorithm selection which they impact.}
     \label{analysis_process}
 \end{center}
\end{figure}
}

\subsection{Characterizing the Problem}

The framework is a mechanism to characterize an analysis problem and to determine the algorithms that best matches the problem characterization. The three factors each drive analytical approach selection and analytical technique selection. The factor \textit{assigned task} pertains to the problem provided by a decision-maker. The analyst must decipher the intent of the assignment from the lexicon of the decision-maker into specific analytical terms, which are listed under the \textit{Task}. This list of terms, called \textit{considerations} is shown in Figure \ref{factors/considerations} for each factor.  The considerations for the factor \textit{data} describe the different formats analysts commonly receive data for analysis problems. The \textit{data} factor is important because it relates to the problem's compatibility with the mathematical mechanics of the analysis technique. Likewise, the considerations for the \textit{resources} factor help the analyst identify which algorithms are compatible with the available resources. The analyst should refer to Figure \ref{factors/considerations} to evaluate and record the considerations for each factor prior to beginning step 1.

{\begin{figure}[h!]
 \begin{center}
    \includegraphics[width=8.5cm]{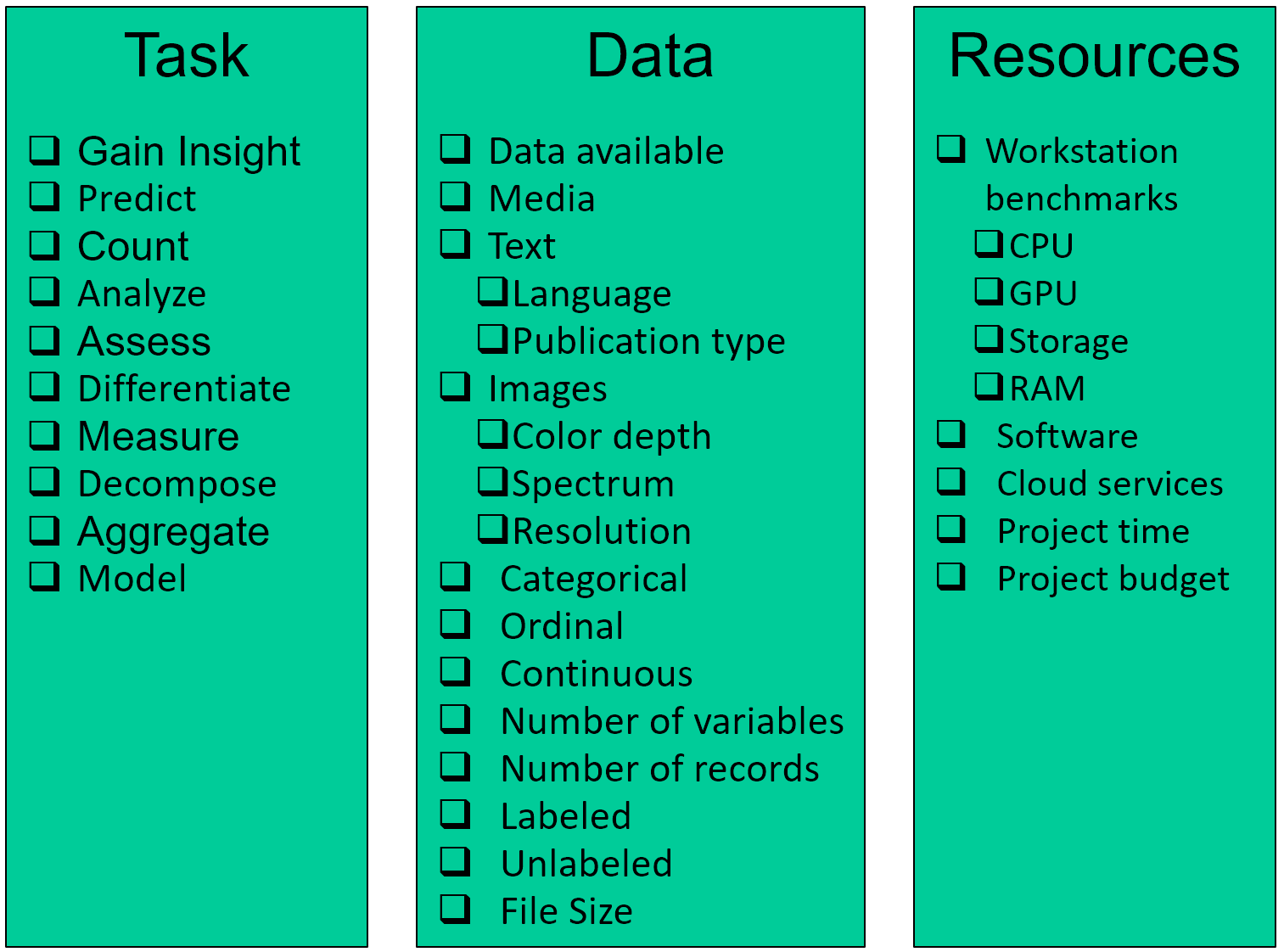}
     \caption{The considerations are shown for each factor which drives analytical approach and analytical technique selection.}
     \label{factors/considerations}
 \end{center}
\end{figure}
}

\subsection{Step 1: Map Problem to Category and Approach}

Step 1 leverages information from the \textit{problem characterization} to identify the appropriate \textit{analytical approaches}. Each \textit{consideration} selected from the \textit{assigned task} factor maps to one or more \textit{categories of analysis}. The \textit{categories of analysis} describe the general goal of the analysis problem \cite{informs2018}. Each \textit{category of analysis} can be implemented by certain \textit{analytical approaches}. The \textit{analytical approach} a technique class referring to the specific type of response the techniques produce. Therefore, the framework leverages a hierarchical taxonomy that groups techniques grouped by both categories of analysis and analytical approaches. Figure \ref{task_to_approach_map} shows the mapping from \textit{assigned task} to \textit{category of analysis}, and the mapping of \textit{category of analysis} to \textit{analytical approach}. 

Since the task of an IDS is to \textit{classify} network users, the \textit{prescriptive} and \textit{predictive} categories of analysis are selected.

{\begin{figure}[h!]
 \begin{center}
    \includegraphics[width=8.5cm]{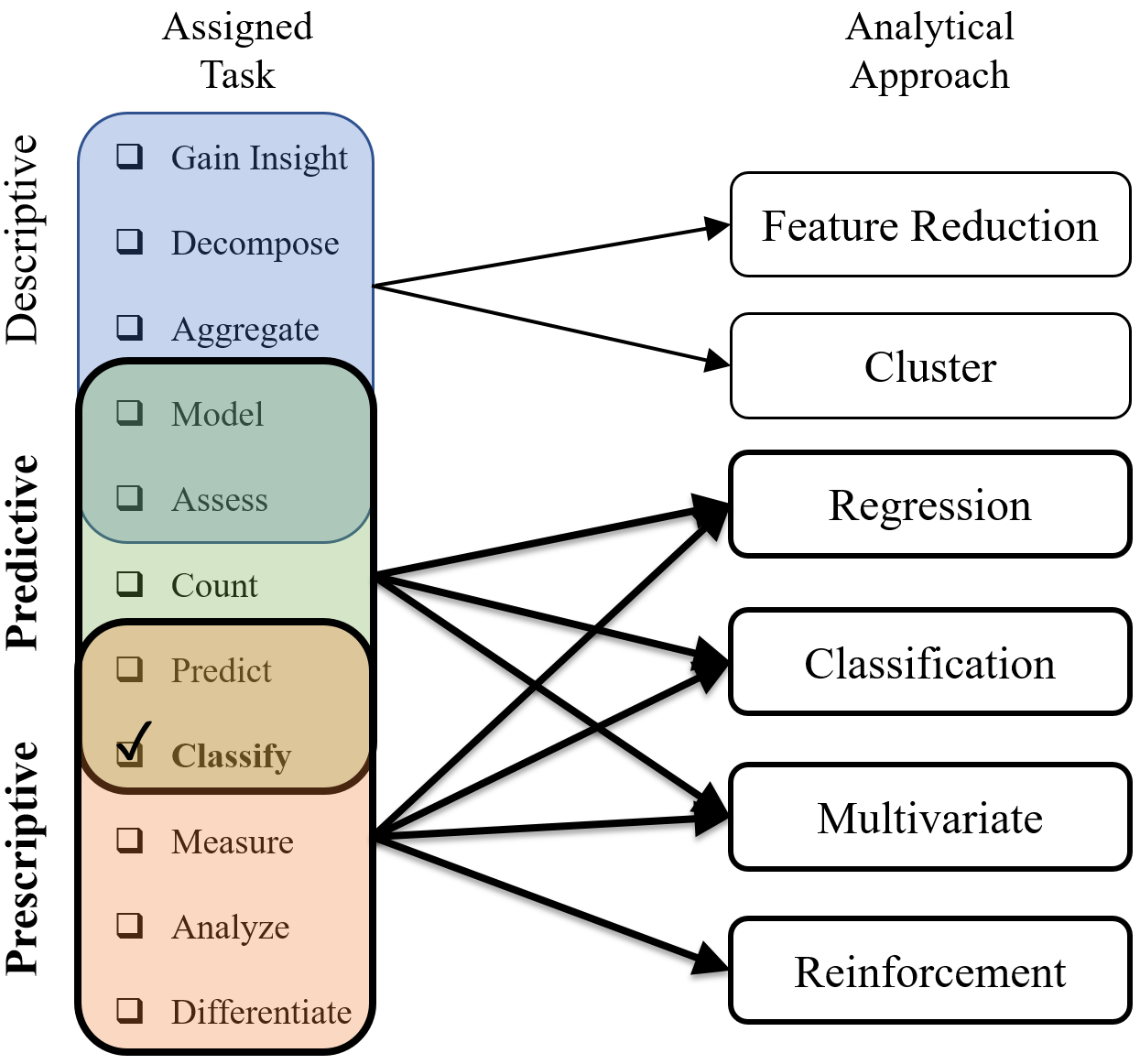}
     \caption{The assigned task for an IDS is classify. Classify is one of 11 common assigned tasks. It belongs to the predictive and prescriptive categories of analysis.}
     \label{task_to_approach_map}
 \end{center}
\end{figure}
}



An excerpt of the proposed taxonomy is presented in Figure \ref{fig:taxonomy}. The taxonomy is built with an object-oriented structure to promote flexibility and expandability. As an example, techniques are shown within the \textit{regression} and \textit{classification} \textit{analytical approaches}. The text \textit{predictive} and \textit{descriptive} appears at the bottom edge of the regression panel to indicate that regression techniques produce results suitable for either of these two \textit{categories of analysis}. The requirements, or required considerations, for each factor are presented with the technique. Compatible training styles are listed to the right of the technique name. The object-oriented structure allows new techniques to be easily added and new attributes to be included.

{\begin{figure}[h!]
 \begin{center}
    \includegraphics[width=8.5cm]{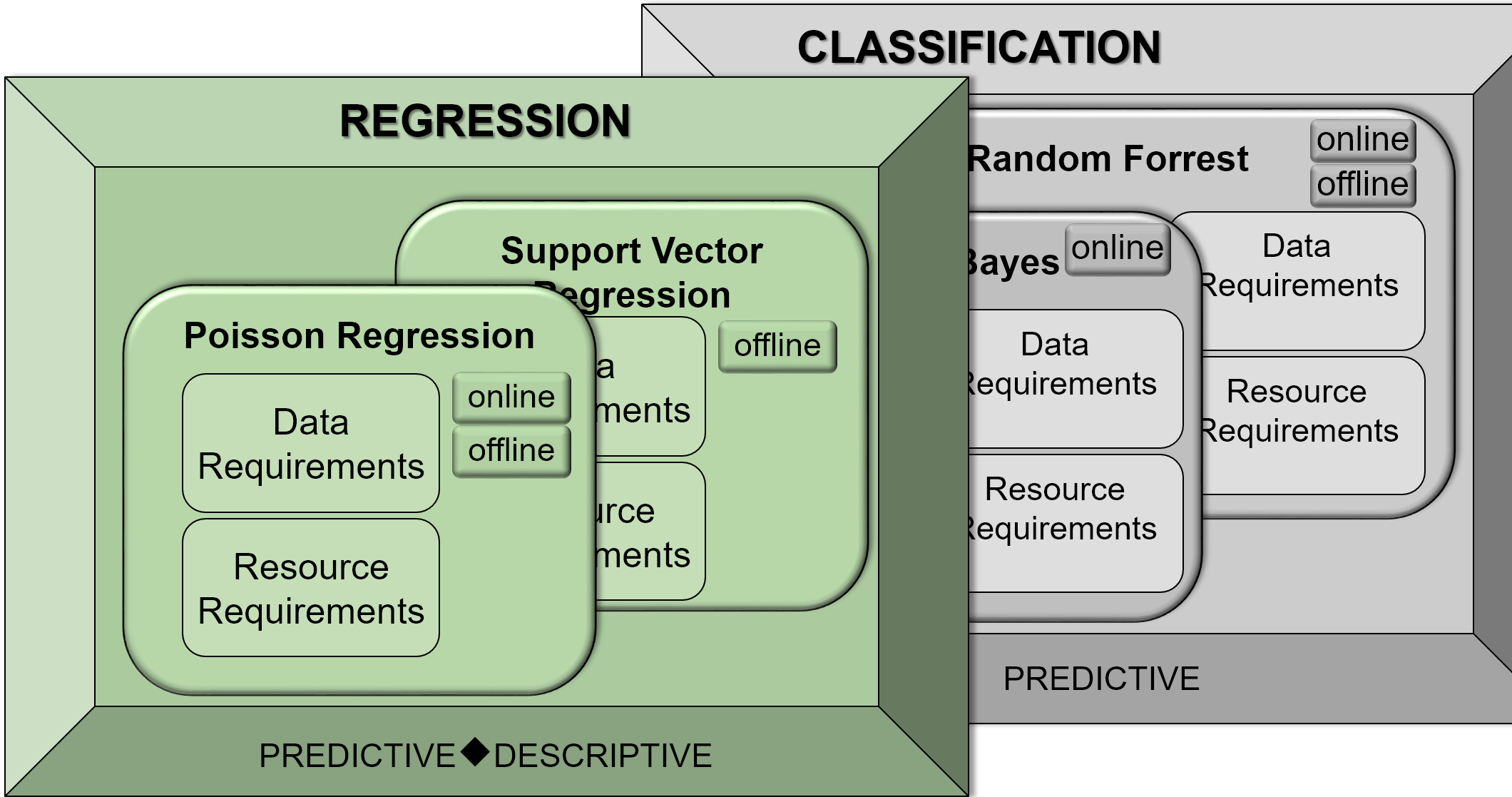}
     \caption{A portion of the proposed taxonomy is high-lighted to show its structure.}
     \label{fig:taxonomy}
 \end{center}
\end{figure}
}

\subsection{Step 2: Rank Techniques}

The framework identifies a subset of techniques that are compatible for the problem according to application. Next, the framework leverages the remaining three factors \textit{data, resources} and \textit{experience} to discern aspects of technique compatibility relating to the mechanics of the mathematical model. Step two predicts the utility scores of each algorithm from these factors according to two strategies: \textit{rules of thumb} and \textit{meta-learning}. They are presented in parallel below. 

\subsubsection{Rules-of-Thumb}

A logical decision tree is used to assign a preference rank among candidate algorithms. The decision tree is built according to rules-of-thumb regarding features of the data. The features pertaining to data also impact the compatibility of techniques in respect to resources. Thus, it is justified to use the same decision tree, Figure \ref{decision_tree}, to adjudicate the scores for both factors.

{\begin{figure}[h!]
 \begin{center}
    \includegraphics[width=8.5cm]{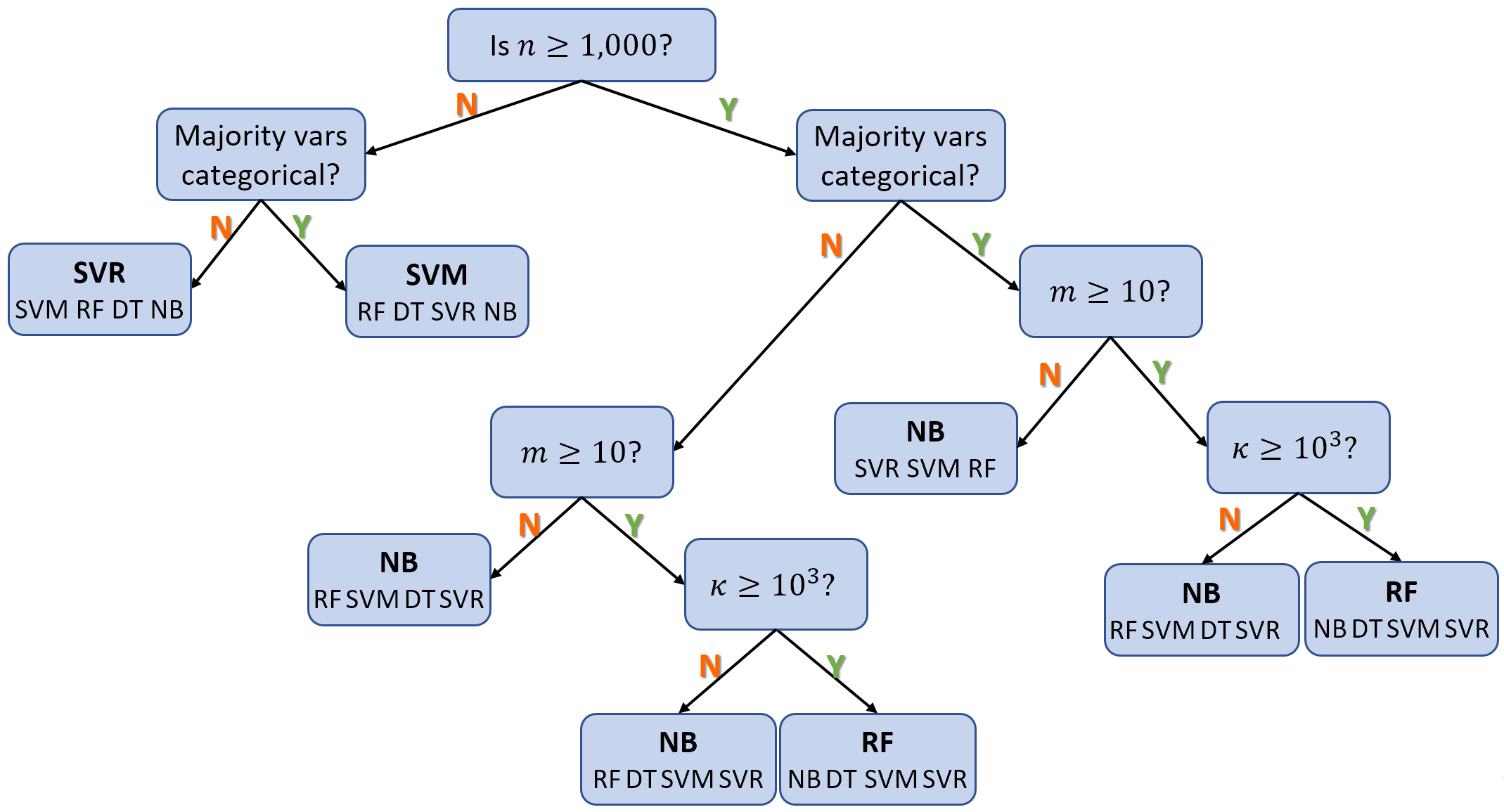}
     \caption{The decision tree represents the logical tests used to rank order the preference of each algorithm via the rules of thumb strategy}.
     \label{decision_tree}
 \end{center}
\end{figure}
}

\subsubsection{Meta-Learning}

A meta-learner is constructed in Python 3.7. All data sets are pre-processed according to best practices for data mining. Base learning is performed on 14 benchmark data sets with 20 repetitions. For each repetition, the data sets were split into training and test sets with an 80/20 ratio and with stratification. The KDDTest+ and KDDTrain+ sets were obtained having already been split into a master test set and a training set. 12 meta-features of each data set were stored as predictor data for the meta-learner; the mean observed recall was stored as the target. The meta-learner was trained to model recall as a function of meta-features using a support vector regression algorithm. The radial basis function kernel was selected. The regularization parameter was set at 1.0, and the kernel coefficient was auto-scaled as a function of the number of features and predictor variance. All other parameters followed Scikit-learn defaults \cite{scikit-learn}. Pseudo-code of the meta-learner is presented in Algorithm 1.

\begin{algorithm}
\SetKwData{predrecalldata}{Predicted Recall} \SetKwData{obsrecalldata}{Observed Recall} \SetKwData{data}{Repository}\SetKwData{set}{Dataset}\SetKwData{testset} {Test Dataset} \SetKwData{sets}{Datasets}\SetKwData{Up}{up}\SetKwData{col}{column}
\SetKwFunction{Union}{Union}\SetKwFunction{FindCompress}{FindCompress}
\SetKwInOut{Input}{input}\SetKwInOut{Output}{output}\SetKwFunction{feature}{featureExtract()}\SetKwFunction{trainbase}{trainBaseLearner()}\SetKwFunction{testbase}{TestBaseLearner()} 
\SetKwFunction{trainmeta}{TrainMetaLearner()} \SetKwFunction{predictrecall}{PredictRecall()}
\SetKwFunction{pca}{PCA()} \SetKwFunction{minimax}{miniMax(0,1)} \SetKwFunction{onehot}{OneHot()}
\SetKwData{alg}{Algorithm} \SetKwData{alglist}{Candidate Algorithms}
\SetKwFunction{record}{RecordRecall()}
\SetKwFunction{rank}{RankAlgorithms()}

\Input{\data of \sets, \alglist}
\Output{\predrecalldata \& \obsrecalldata of \testset using \alglist}
\BlankLine
Pre-processing\\
\For{all \set in \data}{
\For{all \col in \set}{\label{forins}

\lIf{\col is numerical}{\\\minimax\\ \pca\\ \minimax}
\lElse{\col is categorical}{\onehot}
}}\BlankLine

Feature Extraction\\
\For{all \set in \data}{
\feature

}\BlankLine

Base Learning\\
\For{all \alg}{
\For{all \set in \data}{
\trainbase\\
\testbase\\
\record\\
\rank}}\BlankLine

Meta Learning\\
\For{all \alg}{
\trainmeta\\
\predictrecall\\
\record\\
\rank}

\caption{Pseudo-code of Meta-learner}\label{algo_disjdecomp}
\end{algorithm}

The candidate algorithms are selected because they are members analytical approaches derived in step 1 of Section 3.2. Clearly, these are not the only algorithms that fall into the applicable analytical approach. Rather, they represent a demonstrative taxonomy. Note that the Scikit-learn default settings are selected on each algorithm to show generality of the algorithm selection framework.

Training data sets are selected from easily accessible benchmark repositories. The first five are selected to provide diversity of meta-features to the meta-learner and improve the robustness of the model. The nine subsets of KDDTrain+ are selected to provide statistical information consistent with the KDDTest+ data set. KDDTrain+ is split into subset to promote diversity of meta-features but also reduce the number of records in the training set which is an order of magnitude greater than the number of records in the test set. A uniform random number generator is used to determine the number of rows allocated to each training set. Rows are not re-arranged during the subset process. 

\begin{enumerate}
    \item Training Data
\begin{enumerate}
    \item Heart: Predict presence of heart disease from 13  predictor variables \cite{heartdata}
    \item Framingham: Predict presence of heart disease in the Framingham study from 15 predictor variables \cite{framdata}
    \item Spam: Predict if an email is spam based on six predictor variables \cite{Spamdata}
    \item Loan: Predict whether a consumer purchases a loan from Thera Bank based on 12 predictor variables \cite{loandata}
    \item Cancer: Predict whether a patient has breast cancer from 30 predictor variables collected in a fine needle aspirate procedure \cite{Cancer_data}
    \item Nine subsets of KDDTrain+: Predict whether a network activity record is normal or malicious from four categorical and 39 numerical predictor variables \cite{NSLKDD_UNB}
\end{enumerate}

\item{Test Data}
    \begin{enumerate}
        \item KDDTest+: Predict whether a network activity record is normal or malicious from four categorical and 39 numerical predictor variables \cite{NSLKDD_UNB}
    \end{enumerate}

\end{enumerate}

The choice of meta-features was adopted from \cite{williams_thesis}. The following meta-features were used as predictor data by the meta-learner to model expected recall.

\begin{itemize}
    \item Number of Rows  
    \item Number of Columns
    \item Rows to Columns Ratio
    \item Number of Discrete Columns
    \item Maximum number of factors among discrete columns
    \item Minimum number of factors among discrete columns
    \item Average number of factors among discrete columns
    \item Number of continuous columns
    \item Gradient average
    \item Gradient minimum
    \item Gradient maximum
    \item Gradient standard deviation
\end{itemize}

\section{Results}

The algorithm selection framework is applied to the task of classifying network traffic as malicious or normal. Problem characterization is performed in step 1 to identify \textit{prescriptive} and \textit{predictive} as the categories of analysis.  This leads to four analytical approaches, namely regression, classification, multivariate, and reinforcement. Five example algorithms which meet this criteria are taken from a notional taxonomy. 

In step 2, candidate algorithms are ranked in order of preference by each recommendation strategy, rules-of-thumb and meta-learner. Both strategies yielded support vector regression as the most highly recommended algorithm. According to the mean observed recall, random forest was the best performing algorithm to detect malicious activity from the KDDTest+ data set. The standard deviations of observed recall on each algorithm were very low. The 90\% Bonferroni confidence interval for support vector machine (SVM) and support vector regression (SVR) overlapped, indicating statistically identical recall performance. All other mean values were statistically unique. Further, the Spearman's coefficient of rank correlation was not statistically significant, largely due to the small size of the rank scheme. Since neither recommendation strategy succeeded in predicting the best performing algorithm, recall efficiency is introduced. Recall efficiency, Equation \ref{recall_eff}, is the ratio of the recall observed by the top recommended algorithm to best observed recall. 

\begin{equation}
\label{recall_eff}
    E_{R}=\frac{R_{best Rec}}{R_{best Obs}}
\end{equation}

The recall efficiency for SVR, the top recommendation of both strategies, is 0.98. Table \ref{tab:results_table} presents a summary of the results including observed mean recall, meta-learner predicted recall, mean runtime, standard deviation of observed recall, observed ranks, rule-of-thumb predicted ranks, and meta-learner predicted ranks. Figure \ref{ML_Wise_Bar} outlines the mean observed recall and the recall predicted by the meta-learner for each algorithm.

\begin{table*}[]
\caption{Results compare the recommendations of each strategy to observed algorithm performance.}
\label{tab:results_table}
\begin{tabular}{@{}llllllll@{}}
\toprule
{\color[HTML]{000000} }              & {\color[HTML]{000000} \begin{tabular}[c]{@{}l@{}}Observed\\  Mean Recall\end{tabular}} & {\color[HTML]{000000} \begin{tabular}[c]{@{}l@{}}Meta-Learner\\ Predicted Recall\end{tabular}} & {\color[HTML]{000000} \begin{tabular}[c]{@{}l@{}}Mean \\ Runtime (s)\end{tabular}} & {\color[HTML]{000000} \begin{tabular}[c]{@{}l@{}}SD of Observed\\ Recall\end{tabular}} & {\color[HTML]{000000} \begin{tabular}[c]{@{}l@{}}Observed \\ Ranks\end{tabular}} & {\color[HTML]{000000} \begin{tabular}[c]{@{}l@{}}Rules-of-Thumb\\ Predicted Ranks\end{tabular}} & {\color[HTML]{000000} \begin{tabular}[c]{@{}l@{}}Meta-Learner \\ Predicted Ranks\end{tabular}} \\ \midrule
{\color[HTML]{000000} Decision Tree} & {\color[HTML]{000000} 0.975340865}                                                     & {\color[HTML]{000000} 0.891227551}                                                             & {\color[HTML]{000000} 0.188214}                                                    & {\color[HTML]{000000} 0.004141}                                                        & {\color[HTML]{000000} 2}                                                         & {\color[HTML]{000000} 4}                                                                        & {\color[HTML]{000000} 5}                                                                       \\
{\color[HTML]{000000} Random Forest} & {\color[HTML]{000000} 0.982216595}                                                     & {\color[HTML]{000000} 0.935765698}                                                             & {\color[HTML]{000000} 1.593553}                                                    & {\color[HTML]{000000} 0.003063}                                                        & {\color[HTML]{000000} \textbf{1}}                                                & {\color[HTML]{000000} 3}                                                                        & {\color[HTML]{000000} 3}                                                                       \\
{\color[HTML]{000000} Naive Bayes}   & {\color[HTML]{000000} 0.863400857}                                                     & {\color[HTML]{000000} 0.952576618}                                                             & {\color[HTML]{000000} 0.007137}                                                    & {\color[HTML]{000000} 0.007961}                                                        & {\color[HTML]{000000} 5}                                                         & {\color[HTML]{000000} 2}                                                                        & {\color[HTML]{000000} 2}                                                                       \\
{\color[HTML]{000000} SVM}           & {\color[HTML]{000000} 0.959329957}                                                     & {\color[HTML]{000000} 0.925262066}                                                             & {\color[HTML]{000000} 8.602973}                                                    & {\color[HTML]{000000} 0.003163}                                                        & {\color[HTML]{000000} 4}                                                         & {\color[HTML]{000000} 5}                                                                        & {\color[HTML]{000000} 4}                                                                       \\
{\color[HTML]{000000} SVR}           & {\color[HTML]{000000} 0.962446436}                                                     & {\color[HTML]{000000} 0.96525973}                                                              & {\color[HTML]{000000} 7.647529}                                                    & {\color[HTML]{000000} 0.003088}                                                        & {\color[HTML]{000000} 3}                                                         & {\color[HTML]{000000} \textbf{1}}                                                               & {\color[HTML]{000000} \textbf{1}}                                                              \\ \bottomrule
\end{tabular}
\end{table*}

{\begin{figure}[h!]
 \begin{center}
    \includegraphics[width=8.5cm]{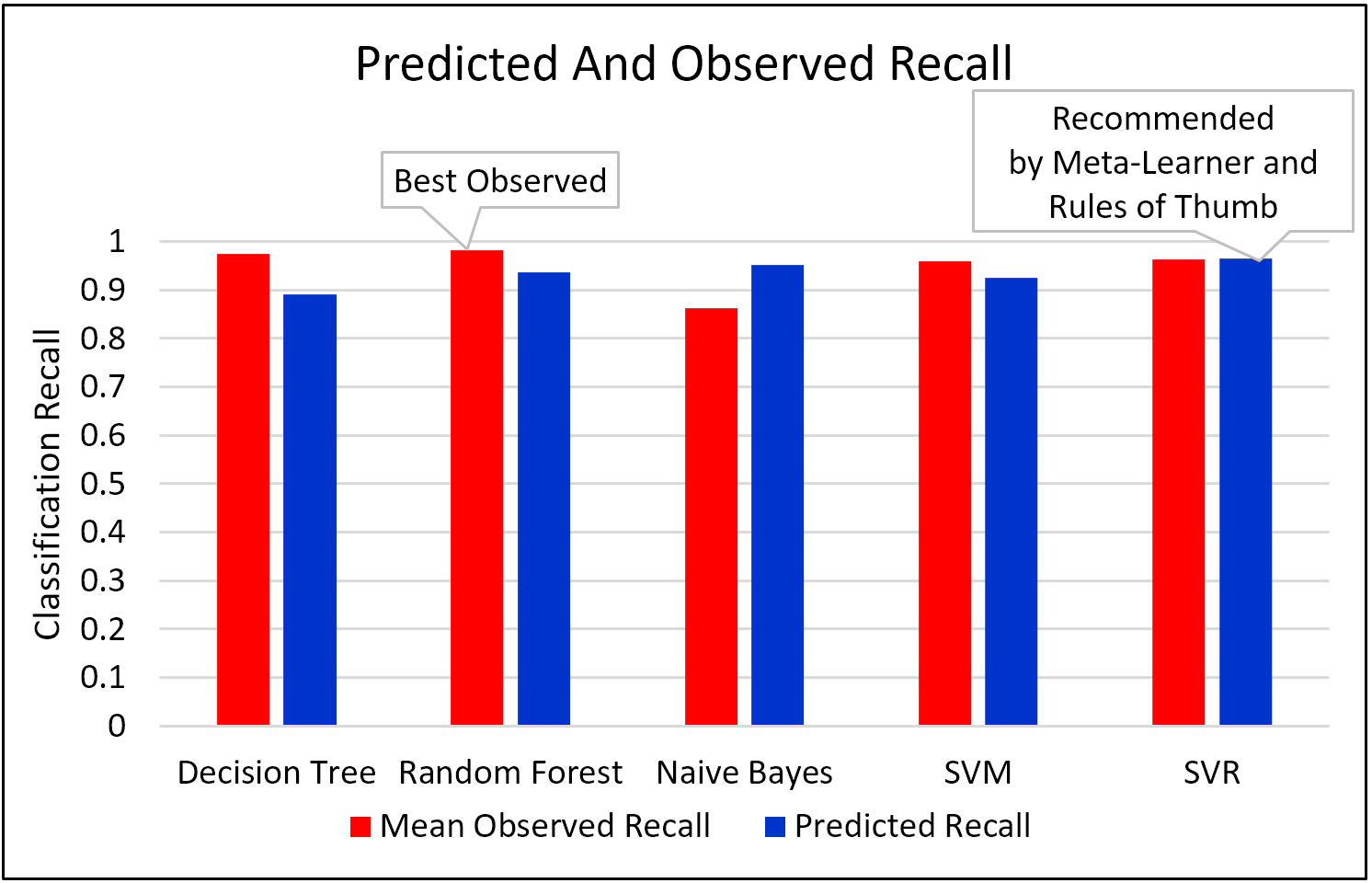}
     \caption{The assigned task for an IDS is classify. Classify is one of 11 common assigned tasks. Classify falls into the predictive and prescriptive categories of analysis.}
     \label{ML_Wise_Bar}
 \end{center}
\end{figure}
}

It is difficult to ascertain whether either of the recommendation strategies employed in this study were successful. While the recall efficiency is very high, the rank correlation was not conclusive. All base learners produced very high and very similar recalls. It would therefore be difficult for any model to discern the true rank preference. The meta-model employed the meta-features according to the precedent set by \cite{williams_thesis}, however there were many more proposed by \cite{Cui_meta} that were not used. Furthermore, the meta-learner was trained by only 14 data sets, significantly less than used in \cite{woods_thesis}. Providing more training sets, especially from the domain of network traffic, would likely improve the predictive capability of the meta-learner.

As a whole, the framework is beneficial even when it does not recommend the true best performer. The framework consistently filters techniques that are incompatible with the problem characterization. Further, the framework identifies five viable options, each of which perform excellently. 

\section{Conclusion}

Cyber attack detection from an IDS using any of the recommended algorithms could reasonably be deemed successful. Neither of the two recommendation strategies demonstrated perfect results. They did, however, show enough promise to motivate further investigations. Fundamentally, the meta-data and user input collected by the framework does contain information capable of consistently predicting a good analysis technique for a problem. Notably, there were algorithms from distinct analytical approaches that performed well on the same task. The process of problem characterization fits well into the framework but does require further refining.  The rule-of-thumb decision tree provided intelligible recommendation logic whereas the meta-learner is a black box model. Future work should use the Gini criterion to optimize the decision tree. Further, the meta-learner should be improved to include more meta-features and training sets. There is a close connectedness in having a useful taxonomy of algorithms and a successful algorithm selection. This relationship is only beginning to be understood. Wireless IDS already provide good classification performance, however, algorithm selection, hyper-parameter tuning and feature engineering suffer from the time-costly trial-and-error practice. The algorithm selection framework may be a step forward in reducing this cost. 

\bibliographystyle{ACM-Reference-Format}
\bibliography{sample-base.bib}

\end{document}